\documentclass[aps,prc,notitlepage,twocolumn,floatfix,nofootinbib,superscriptaddress,showkeys,10pt]{revtex4-2}
    
\usepackage[utf8]{inputenc}          
\usepackage{graphicx}         
\usepackage[table]{xcolor}
\usepackage{array,dcolumn,longtable} 
\usepackage{amsmath,amssymb,amsfonts,slashed} 
\usepackage{mathtools}          

\usepackage[linktocpage,breaklinks]{hyperref}

\usepackage{txfonts}
\usepackage{bm}
\usepackage{stmaryrd}
\usepackage{tensor}
\usepackage[utf8]{inputenc}

\usepackage{epsfig}
\usepackage{epstopdf}

\usepackage{natbib}
\usepackage{cleveref}

\usepackage{setspace}

\def\l{\left}
\def\r{\right}
\newcommand{\be}{\begin{equation}}
\newcommand{\ee}{\end{equation}}
\newcommand{\bea}{\begin{eqnarray}}
\newcommand{\eea}{\end{eqnarray}}
\newcommand{\bml}{\begin{subequations}}
\newcommand{\eml}{\end{subequations}}

\definecolor{purple}{rgb}{0.52, 0., 0.52}

\begin{document}

\title{Causality violations in realistic simulations of heavy-ion collisions}

\author{Christopher Plumberg}
\affiliation{Illinois Center for Advanced Studies of the Universe, Department of Physics, University of Illinois at Urbana-Champaign, Urbana, IL 61801, USA}
\author{Dekrayat Almaalol}
\affiliation{Illinois Center for Advanced Studies of the Universe, Department of Physics, University of Illinois at Urbana-Champaign, Urbana, IL 61801, USA}
\affiliation{Department of Physics, Kent State University, Kent, OH 44242, USA}
\author{Travis Dore}
\affiliation{Illinois Center for Advanced Studies of the Universe, Department of Physics, University of Illinois at Urbana-Champaign, Urbana, IL 61801, USA}
\author{Jorge Noronha}
\affiliation{Illinois Center for Advanced Studies of the Universe, Department of Physics, University of Illinois at Urbana-Champaign, Urbana, IL 61801, USA}
\author{Jacquelyn Noronha-Hostler}
\affiliation{Illinois Center for Advanced Studies of the Universe, Department of Physics, University of Illinois at Urbana-Champaign, Urbana, IL 61801, USA}

\date{\today}

\begin{abstract}
Causality is violated in the early stages of state-of-the-art heavy-ion hydrodynamic simulations. Such violations are present in up to 75\% of the fluid cells in the initial time and only after 2-3 fm/c of evolution do we find that $50\%$ of the fluid cells are definitely causal. Superluminal propagation reaches up to 15\% the speed of light in some of the fluid cells. The inclusion of pre-equilibrium evolution significantly reduces the number of acausal cells. Our findings suggests that relativistic causality may place constraints on the available parameter space of heavy-ion collision simulations when factored into more thorough statistical analyses.
\end{abstract}
\maketitle

\noindent\underline{\textit{Introduction:}}
Relativistic viscous hydrodynamics is vital for the phenomenological modeling of ultrarelativistic heavy ion-collisions \cite{Heinz:2013th,deSouza:2015ena,Romatschke:2017ejr}. Confirmed predictions at the LHC \cite{Noronha-Hostler:2015uye, Niemi:2015voa, Adam:2015ptt} at the percent level, and the ability to fit standard observables \cite{Song:2010mg, Bozek:2012qs, Gardim:2012yp, Bozek:2013uha, Niemi:2015qia, Ryu:2015vwa, McDonald:2016vlt, Bernhard:2016tnd, Gardim:2016nrr, Alba:2017hhe, Giacalone:2017dud, Eskola:2017bup, Weller:2017tsr, Schenke:2019ruo}, provide strong evidence for the formation of a fluid-like state of matter known as the quark-gluon plasma (QGP) in high-energy nuclear collisions \cite{Shuryak:2014zxa}. 

Comparisons to experimental data require modeling all the stages of a heavy-ion collision: the initial \cite{Schenke:2012wb,Moreland:2014oya,Niemi:2015qia}, the pre-equilibrium stages \cite{Xu:2004mz,Broniowski:2008qk,Liu:2015nwa,Kurkela:2018wud,Kurkela:2018vqr},  relativistic hydrodynamics \cite{Romatschke:2017ejr}, and hadronic interactions \cite{Bass:1998ca,Bleicher:1999xi,Nara:1999dz,Lin:2004en,Weil:2016zrk}. Relativistic viscous fluid-dynamics is currently determined by equations of motion \cite{Israel:1979wp,Baier:2007ix,Denicol:2012cn} for an extended set of dynamical variables which include the temperature, chemical potentials, and flow velocity as well as  non-equilibrium currents, such as the shear-stress tensor, $\pi_{\mu\nu}$, the bulk scalar, $\Pi$, and diffusion currents \footnote{Viable descriptions of relativistic viscous fluids can also be obtained at first-order in derivatives using only the hydrodynamic variables, see \cite{Bemfica:2017wps,Kovtun:2019hdm,Bemfica:2019knx,Hoult:2020eho,Bemfica:2020zjp}}. These simulations have provided key insight into the temperature dependence of the QGP's transport coefficients  \cite{Bernhard:2016tnd,Moreland:2018gsh, Bernhard:2019bmu,Auvinen:2020mpc,Everett:2020xug,Nijs:2020roc}. 

The applicability of hydrodynamics to small and short-lived nuclear systems is far from trivial. Very large initial spatial gradients  occur \cite{Bjorken:1982qr,Schenke:2012wb,Niemi:2014wta,Noronha-Hostler:2015coa}, driving the system far-from-equilibrium. Furthermore, collective behavior compatible with hydrodynamics was found in even smaller systems (e.g. pA collisions) 
\cite{Chatrchyan:2013nka,ABELEV:2013wsa,Aad:2013fja,PHENIX:2018lia}.  While progress on understanding far-from-equilibrium relativistic hydrodynamics  has been made \cite{Heller:2015dha,Florkowski:2017olj,Berges:2020fwq}), traditionally \cite{landaulifshitzfluids} hydrodynamics is only expected to accurately describe the long time, long wavelength behavior of systems close to equilibrium. 

A strong connection exists between the initial energy density's spatial anisotropy and the final flow harmonics  \cite{Teaney:2010vd,Gardim:2011xv,Niemi:2012aj,Teaney:2012ke,Qiu:2011iv,Luzum:2013yya,Gardim:2014tya,Betz:2016ayq} that begins to break down in small systems  \cite{Noronha-Hostler:2015coa,Mazeliauskas:2015vea,Sievert:2019zjr,Zhao:2020pty} due to significant initial out-of-equilibrium contributions \cite{Schenke:2019pmk}. Thus, the emergence of hydrodynamics and its domain of applicability have direct relevance to QGP phenomenology.

In the far-from-equilibrium domain, dissipative contributions to the energy-momentum tensor of the system can become comparable to the equilibrium pressure $P$. Then, viscous terms contribute significantly to the fluid evolution \cite{HISCOCK1988509} and constraints on the transport coefficients derived in \cite{Hiscock:1983zz,Olson:1989ey,Pu:2009fj} using linearized perturbations around equilibrium are insufficient to ensure a well-defined causal evolution. 
Current heavy-ion simulations employ transport coefficients to satisfy these linear constraints, but it is unknown whether causality actually holds in such simulations in the nonlinear far-from-equilibrium regime. 

This question can be answered using the new constraints  \cite{Bemfica:2020xym} involving the magnitude of the viscous currents and transport coefficients, which ensure that causality \cite{Hawking:1973uf} holds in the nonlinear regime of the class of 2nd order hydrodynamic equations of motion \cite{Israel:1979wp,Baier:2007ix,Denicol:2012cn} used in heavy-ion simulations.
These constraints define the physically allowable space of out-of-equilibrium corrections to the initial state, providing new theoretical guidance for relativistic viscous hydrodynamics.

In this work we investigate these causality constraints for the most well-behaved scenario in heavy ions collisions simulations: central LHC Pb+Pb collisions. Two state-of-the-art open-source frameworks are used in our study: The first \cite{Shen:2014vra, Moreland:2018gsh, Bernhard:2019bmu, Everett:2020xug} couples T$_{\rm R}$ENTo+free-streaming+VISHNU  and the second framework \cite{Schenke:2010nt, Schenke:2012hg, Schenke:2012wb, Gale:2012rq, Gale:2013da, Kurkela:2018vqr, Kurkela:2018wud} couples IP-Glasma+(K${\o}$MP${\o}$ST)+MUSIC. For compactness, we at times refer to these two frameworks as the ``TFV" and ``IKM" frameworks, respectively, in the text. Both frameworks generically yield causality violations throughout a significant portion of the early time evolution, for typical parameter settings determined via comparisons to experimental data. In the IKM framework, we study if variations in the pre-hydrodynamic phase \cite{Kurkela:2018vqr, Kurkela:2018wud} can ameliorate these violations. Pre-equilibrium evolution significantly reduces acausal behavior, but does not eliminate it. Our analysis suggests that the nonlinear constraints imposed by causality should be taken into account in the assessment of viable regions of hydrodynamic parameter space and, ultimately, in the quantitative extraction of QGP properties.

\begin{figure*}
    \centering
    \includegraphics[keepaspectratio, width=\textwidth]{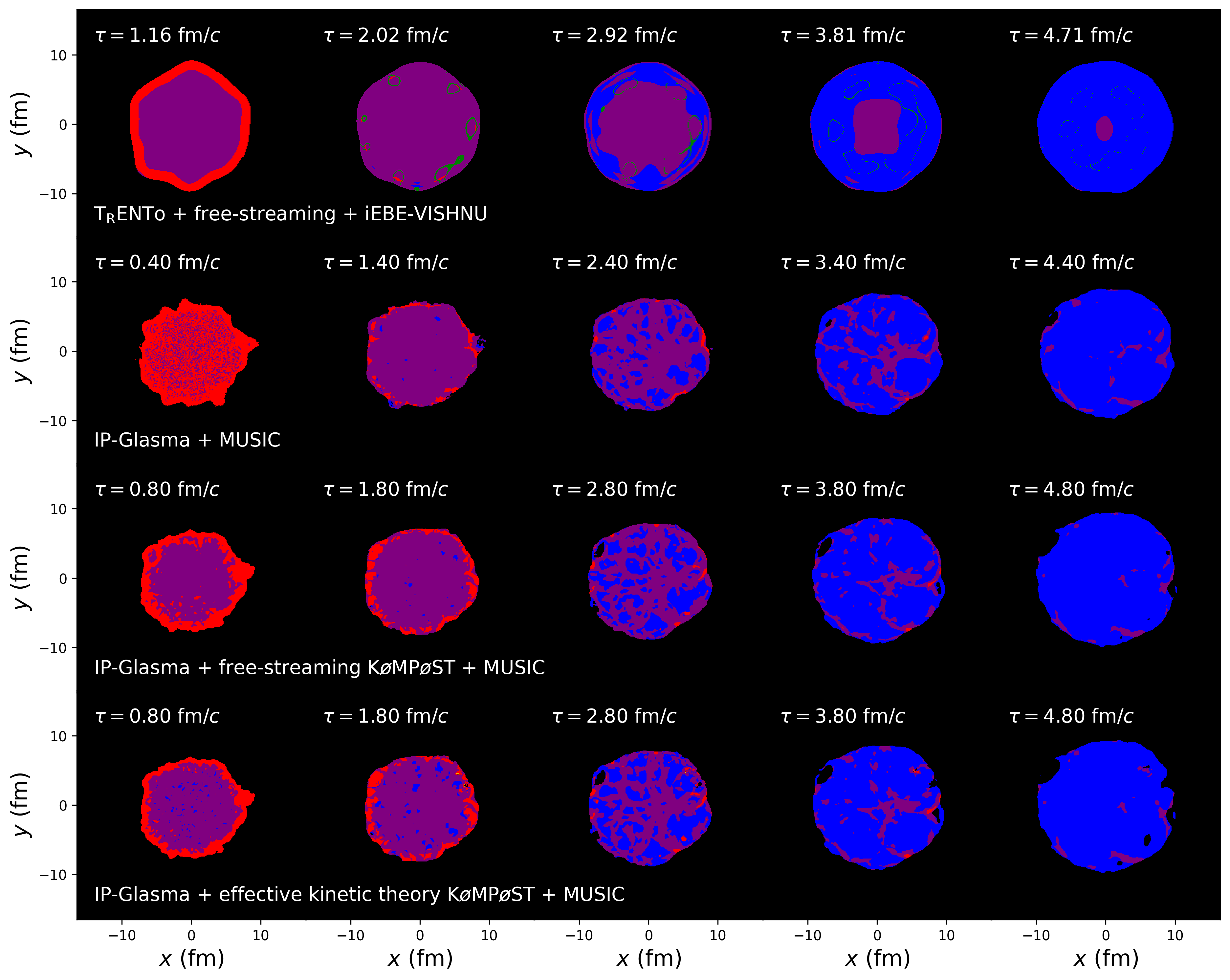}
    \caption{From top to bottom: the TFV scenario and the three IKM scenarios (no K{\o}MP{\o}ST, free-streaming K{\o}MP{\o}ST, EKT K{\o}MP{\o}ST).  Colors correspond to the following cell classifications: causal(blue), acausal (red), and purple (indeterminate).  Cells where the causality analysis is inapplicable are colored green or orange, as discussed in the text.}
    \label{Fig1}
\end{figure*}

\noindent\underline{\textit{Modeling:}} Both frameworks incorporate a fully initialized energy-momentum tensor $T^{\mu\nu}$ in their initial state, have constrained  parameters through a Bayesian analysis \cite{Paquet:2017mny,Moreland:2018gsh, Bernhard:2019bmu}, and have been extensively compared to experimental data. An alternative approach also exists that only initializes the energy density profiles \cite{Alba:2017hhe,Niemi:2015qia}, which  produces relatively equivalent results to experimental data except for a handful of observables \cite{NunesdaSilva:2020bfs,Schenke:2020uqq,Giacalone:2020dln,Giacalone:2020byk,ATLAS:2021kty}. While the  T$_{\rm R}$ENTo and IP-Glasma inital state models have  comparable energy density eccentricities \cite{Moreland:2014oya,Bernhard:2016tnd,Giacalone:2017uqx}, subtle differences remain that are  likely due to the scaling of the initial energy density distribution with the thickness functions \cite{Nagle:2018ybc,Carzon:2020ohg}. Our simulations are performed at zero baryon chemical potential.

In the TFV framework, we adopt the Bayesian tune to LHC p+Pb and Pb+Pb data \cite{Moreland:2018gsh, Bernhard:2019bmu} which combines  T$_{\rm R}$ENTo initial conditions \cite{Moreland:2014oya} with a conformal, pre-hydrodynamic free-streaming phase \cite{Broniowski:2008qk, Liu:2015nwa}, a boost-invariant hydrodynamic phase \cite{Song:2007ux, Shen:2014vra}, and a hadronic afterburner UrQMD \cite{Bass:1998ca, Bleicher:1999xi}. We use the maximum-likelihood parameters  \cite{Bernhard:2019bmu} for the transport coefficients. A single, $\sqrt{s_{NN}} = 2.76$ TeV central Pb+Pb event is generated and the random seed is set to 1  to ensure reproducibility of our results.    The energy density freeze-out criterion is imposed at  $\varepsilon_{FO} \approx 0.265$ GeV/fm$^3$.

In the IKM framework, the initial conditions are from IP-Glasma, coupled to classical Yang-Mills evolution \cite{Schenke:2012hg, Schenke:2012wb}, followed by a boost-invariant hydrodynamics (MUSIC) starting at $\tau=0.4$ fm$/c$ \cite{Gale:2012rq}.  We consider an intervening pre-hydrodynamic phase starting at $\tau=0.1$ fm$/c$ and propagated until $\tau=0.8$ fm$/c$  using K${\o}$MP${\o}$ST \cite{Kurkela:2018vqr, Kurkela:2018wud}: ``FS"  free-streaming or ``EKT"  effective kinetic theory.  Three different scenarios are considered: (i) IP-Glasma + MUSIC; (ii) IP-Glasma + K${\o}$MP${\o}$ST (FS) + MUSIC; (iii) IP-Glasma + K${\o}$MP${\o}$ST (EKT) + MUSIC. A single, $\sqrt{s_{NN}} = 2.76$ TeV central Pb+Pb event is generated with a random seed of 1615404198. All scenarios use  $\eta/s = 0.12$ and the $[\zeta/s](T)$ parameterization from \cite{Schenke:2020mbo}, and freeze-out occurs at $\varepsilon_{FO} = 0.18$ GeV/fm$^3$ or at $T_{FO}= 145 $ MeV. The pressure $P$ is from the lattice QCD-based equations of state in both frameworks \cite{Bazavov:2014pvz, Moreland:2015dvc}.

\begin{figure*}
    \centering
    \includegraphics[keepaspectratio, width=\linewidth]{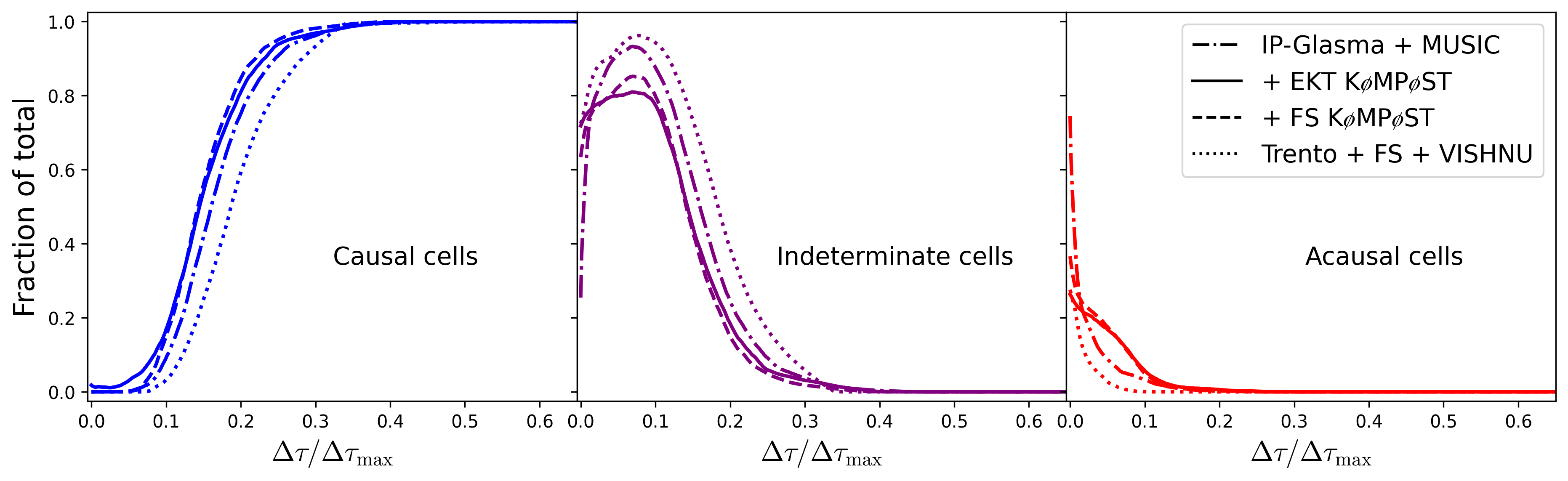}
    \caption{Fractions of the number of hydrodynamic cells ($\varepsilon \geq \varepsilon_{FO}$) that are causal (left), indeterminate (center), or acausal (right)  vs. the rescaled time evolution in each framework.}
    \label{Fig2}
\end{figure*}

The constraints  from \cite{Bemfica:2020xym} apply to the  Israel-Stewart-like \cite{Israel:1979wp,Baier:2007ix,Denicol:2012cn} equations of motion used in both frameworks. They were found by determining the characteristic velocities (i.e., the propagation modes) of the corresponding nonlinear system of PDEs, which were used to obtain a set of \emph{necessary} conditions for causality, i.e., the system must satisfy these conditions to be causal. \emph{Sufficient} conditions for causality indicate that causality is guaranteed to hold. Both sets of conditions correspond to simple inequalities involving  transport coefficients and  viscous currents, i.e. $\Pi$ and the four eigenvalues $\{0,\Lambda_i\}$ of $\pi^\mu_\nu$ (with $i=1,2,3$ and $\sum_{i=1}^3\Lambda_i=0$), which can be evaluated at each time step. For the explicit expressions of the constraints, see \cite{Bemfica:2020xym} or the Supplemental Material \cite{SupplementaryMaterial}.

We sort grid points in the simulations into three different categories, identified by  colors: Blue: points at which the sufficient conditions (and consequently the necessary conditions) hold, hence causality is respected. Red: points at which one or more necessary conditions (and consequently sufficient conditions) are violated, hence causality is unquestionably violated. Purple: points at which all necessary conditions are satisfied but one or more sufficient conditions fail, hence the analysis cannot determine if causality is violated. On very rare occasions, points occur where the pre-conditions \cite{Bemfica:2020xym} for the applicability of the causality analysis fail to hold.  Here, this is typically due, e.g., to values of $\Lambda_i$ for which $\varepsilon + P + \Pi + \Lambda_i$ is not positive. We color these points orange in our plots below.  Green points denote the case where the diagonalization of $\pi^\mu_\nu$ fails and $\pi^{\mu\nu}u_\mu\neq 0$.  However, both orange and green points occur so infrequently that they are barely visible in the plots and will be neglected in the following.

\noindent\underline{\textit{Results:}} The time evolution of the causality analysis for a typical Pb+Pb event is shown for the TFV framework and all scenarios of the IKM framework in Fig.~\ref{Fig1}. Note that only the fluid cells that have not yet frozen out are plotted.  The hydrodynamic simulations are all characterized by pervasive violations of causality, particularly in the first 1-2 fm$/c$ of the collision. For the TFV framework, most of the severe causality violation occurs near the edge of the system where Knudsen and inverse Reynolds numbers \cite{Niemi:2014wta,Noronha-Hostler:2015coa} become large, though still above freeze out. For the IKM framework, without K${\o}$MP${\o}$ST, approximately $75\%$ of cells in the initial state violate causality. However, the inclusion of K${\o}$MP${\o}$ST pre-equilibrium evolution significantly reduces the causality violation present in the IP-Glasma initial state,  bringing it down to approximately 1/3 of fluid cells. EKT has a slight improvement over FS but the difference is small.

These plots demonstrate some qualitative features that are likely due to different choices in the transport coefficients in the two frameworks.  For instance, the TFV framework appears to switch the regions at the edge from acausal to causal first and work its way inwards (with a small region of indeterminable cells at the center at late times). In contrast, the IKM scenarios have acausal and indeterminate regions at the edges throughout the expansion but appears to have more causal regions at the center. This may be due to the larger bulk viscosity used in the IKM framework \cite{Schenke:2020mbo, Gale:2020xlg} or to the smoother initial conditions from T$_{\rm R}$ENTo. 

In Fig.~\ref{Fig2} we show the time evolution of the fraction of fluid cells (with $\varepsilon \geq \varepsilon_{FO}$), plotted as a function of the rescaled time $\Delta \tau \equiv \tau - \tau_{\mathrm{hydro}}$ (where $\tau_{\mathrm{hydro}}$ is the time at which hydrodynamics begins).  During roughly the first 20\% of the evolution most of the system's fluid cells are either acausal (red) or indeterminate (purple/green).  All simulations considered do eventually converge to a regime where the hydrodynamic evolution is completely causal everywhere.  These observations hold quite generally for the different events, centralities, and collision systems we considered.  Our results appear to be consistent with those from a recent work \cite{Cheng:2021tnq}, which has also studied causality in AA collisions. However, we find that the fraction of causality-violating cells at a given time can include up to $75\%$ of the system \footnote{We note that the acausal cells account for only about $1-2\%$ of the total number of cells in the simulation throughout the evolution of the system (because the acausal cells turn causal at larger times), which is consistent with \cite{Cheng:2021tnq}.}.

\begin{figure*}
    \centering
    \includegraphics[keepaspectratio, width=\linewidth]{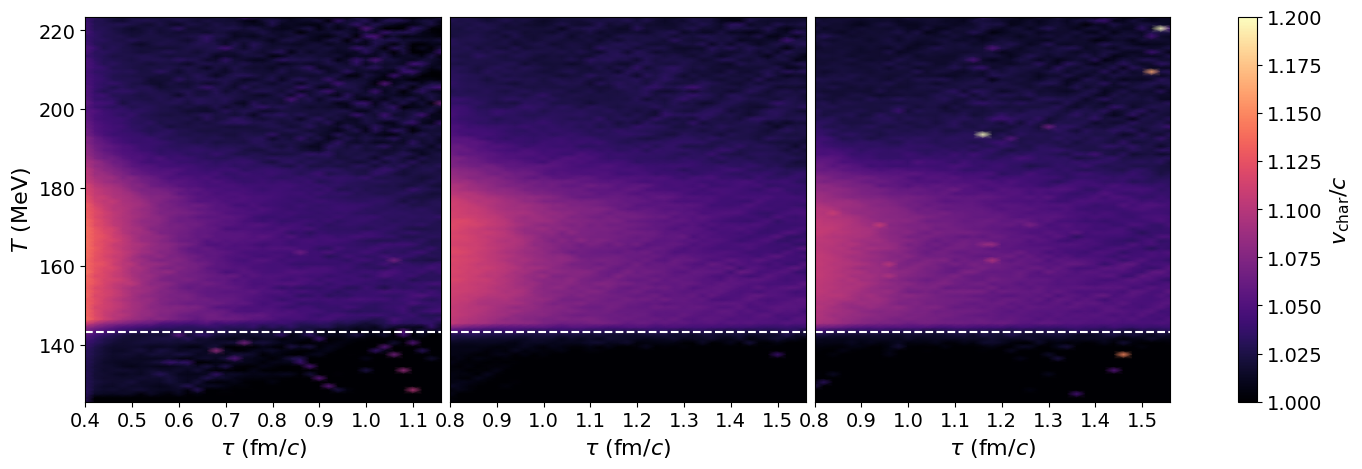}
    \caption{Characteristic velocities for PbPb collision in the IKM framework: Left: No K${\o}$MP${\o}$ST, Middle: Free Streaming, Right: K${\o}$MP${\o}$ST EKT. }
    \label{density_plots}
\end{figure*}
In Fig.\ \ref{density_plots}, we show the times and temperatures over which characteristic velocities were found to propagate faster than the speed of light in the three cases of the IKM framework. The details concerning the calculation of these velocities are given in the Supplemental Material \cite{SupplementaryMaterial} and can also be found in \cite{Bemfica:2020xym}. These show that the characteristic velocities were calculated to be around 15\% greater than the speed of light, most pervasive both at early times and near the transition temperature. One can note that while the calculated super-luminal speeds seem to decrease as pre-equilibrium is turned on, its pervasiveness in time is increased. Also, we note that   the scenario with K${\o}$MP${\o}$ST EKT has small regions of very high superluminal characteristic velocities.

\begin{figure}
    \centering
    \includegraphics[keepaspectratio, width=\linewidth]{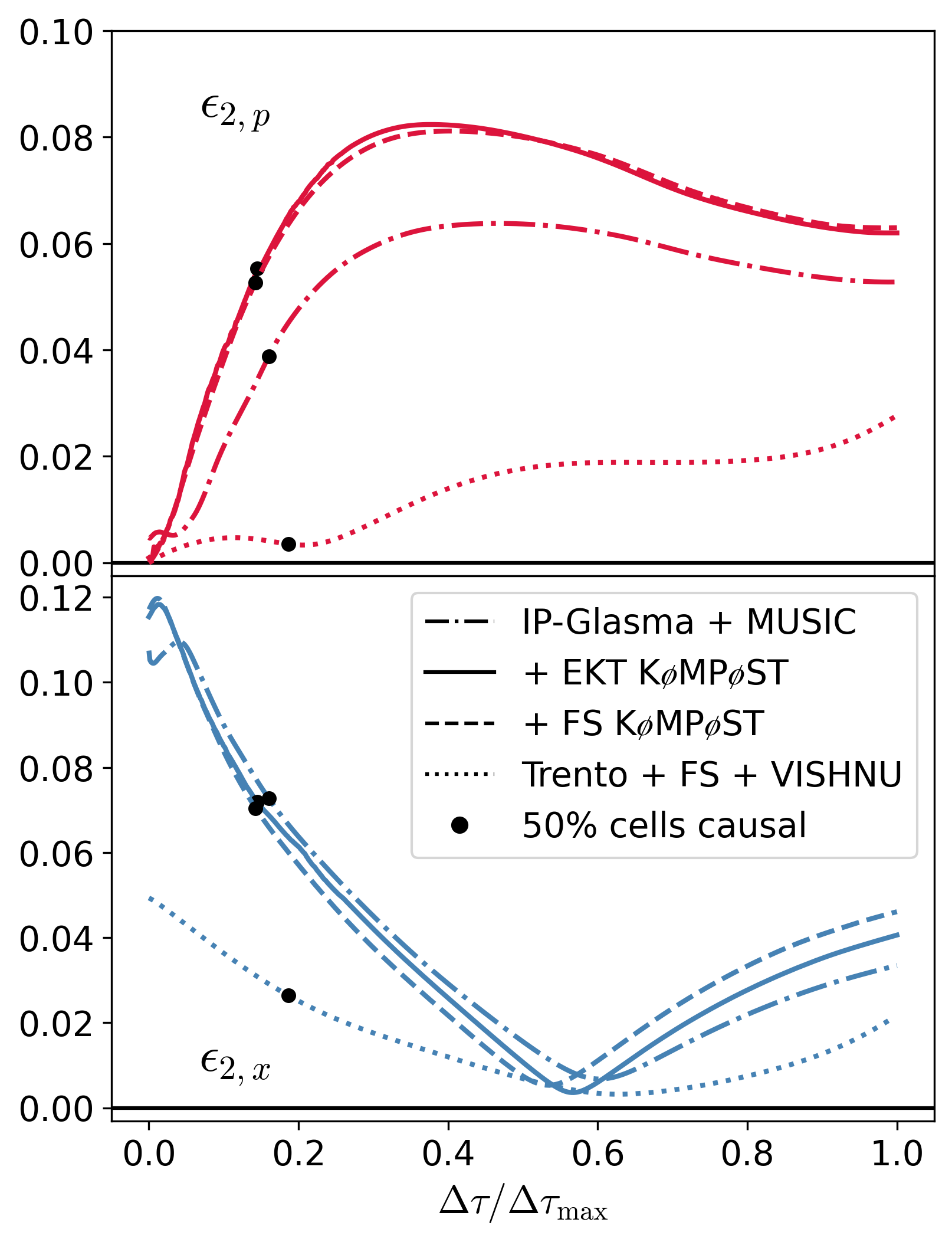}
    \caption{The momentum anisotropy $\epsilon_{2,p}$ (top) and spatial eccentricity $\epsilon_{2,x}$ (bottom) versus time.  Black dots represent the points at which exactly half of the hydrodynamic cells (with $\varepsilon \geq \varepsilon_{FO}$) become explicitly causal.}
    \label{Fig3}
\end{figure}

Further work is needed to explore the consequences of causality violations for experimental observables such as anisotropic flow or the HBT radii \cite{Adamova:2017opl, Acharya:2018lmh}.  In this first study, in lieu of these standard observables, we consider instead the momentum anisotropy $\epsilon_{2,p}$ and the spatial eccentricity $\epsilon_{2,x}$ \footnote{$ \epsilon_{2,p}=\sqrt{\l( \l< T^{xx}-T^{yy}\r>_1^2 + \l< 2T^{xy} \r>_1^2 \r) / \l< T^{xx}+T^{yy} \r>_1^2} \label{e2pdef},\\ \epsilon_{2,x}= \sqrt{\l( \l< x^2-y^2\r>_{e \gamma}^2 + \l< 2 x y \r>_{e \gamma}^2 \r) / \l< x^2+y^2 \r>_{e \gamma}^2},\\ \label{e2xdef} \mathrm{where}\,  \l< f(x,y) \r>_w= \l.\int dx\, dy\, w(x,y) f(x,y) \r/ \int dx\, dy\, w(x,y) \\ \mathrm{and} \hspace{0.3cm}\gamma = \sqrt{1-u_x^2-u_y^2} $}.  We study in Fig.\ \ref{Fig3} how these quantities evolve with time in different scenarios. The black dots indicate the point in time for each scenario when half of the fluid cells are certainly causal. Whereas most of the final $\epsilon_{2,p}$ in the TFV framework is built up after the majority of the system has become causal, in the IKM scenarios the majority of the $\epsilon_{2,p}$ anisotropy is built up at early times (up to $20-30\%$ of evolution time) and nearly half of the final anisotropy is built up when most of the system either explicitly violates causality or the sufficient conditions are not met.  This shows that enforcing causality criteria may lead to measurable effects for final state observables \cite{Cheng:2021tnq}, which should be considered in Bayesian analyses that seek to realistically extract QGP properties. 

Omitting the acausal and indeterminate cells from the calculation leads to significantly different estimates for the initial and final values of $\epsilon_{2,p}$ and $\epsilon_{2,x}$ and, thus, substantially different interpretations of the underlying physics. Generally, the eccentricities are larger when only causal cells are considered (and, conversely, their radii are smaller).  Although we cannot at this stage remove the effects of causality violations entirely from our simulations, these results suggest that both collective dynamics and spatial geometry will be affected once causality constraints are taken into account.

\noindent\underline{\textit{Conclusion:}} In this paper we conclusively showed that there are sizable causality violations in state-of-the-art simulations of heavy-ion collisions. The TFV and IKM frameworks, with parameters constrained by experimental data, yield up to 75\% of fluid cells explicitly violating causality in the earliest stages of central Pb+Pb collisions at the LHC.  Retaining causality in small systems may be even more problematic than in large system (see also \cite{Cheng:2021tnq}), depending on the pre-equilibrium evolution and the model parameter space favored by data (since Knudsen and Reynolds numbers remain large throughout the entire evolution even for intermediate systems \cite{Summerfield:2021oex}). Our causality analysis of a p+Pb event in the TFV framework can be found in the Supplemental Material \cite{SupplementaryMaterial}.

A pre-equilibrium phase prior to the hydrodynamic evolution significantly reduces the amount of causality violation, though it does not fully eliminate it. Much of this analysis depends on our understanding of the pre-equilibrium phase, which is typically modeled in a conformally invariant manner, whereas  the equation of state of quantum chromodynamics \cite{Borsanyi:2013bia,Bazavov:2014pvz,Borsanyi:2016ksw} used in the hydrodynamic evolution is far from  conformal even at the high temperatures probed at top LHC energies at early times (see \cite{NunesdaSilva:2020bfs} for the consequences of matching a pre-equilibrium conformal phase to a nonconformal hydrodynamic evolution). Further improvements in the pre-equilibrium phase \cite{Martinez:2019jbu,Martinez:2019rlp,Kamata:2020mka,Nijs:2020roc}, going beyond conformal and boost invariance, are needed to fix this acausal behavior found in hydrodynamic simulations of the QGP formed in heavy-ion collisions.

Another possible solution would be the systematic implementation of  causality constraints into Bayesian analyses (e.g., \cite{Bernhard:2019bmu}), which would allow the causality requirements to dictate which regions of parameter space are most viable. Such an analysis, performed taking into account both AA and small systems, would be crucial to determine the values of transport coefficients and the initial viscous currents that are physical and compatible with experimental data. Alternatively, when causality violation is concentrated at the edge of the system, a core-corona approach wherein only fluid cells that are causal are run through hydrodynamics (the core) and all other fluid cells (the corona) are hadronized  \cite{Hirano:2005wx,Aichelin:2008mi,Ahmad:2016ods,Kanakubo:2019ogh} may be more applicable.

\begin{acknowledgments}

\noindent\underline{\textit{Acknowledgements:}} We thank M.~Disconzi for discussions and N.~Cruz Camacho, G.~S.~Denicol, M.~Luzum,  A.~Mazeliauskas, B.~Schenke, and C.~Shen for helpful comments concerning the hydrodynamic simulations. J.N.H, T.D., and C.P. are supported by the US-DOE Nuclear Science Grant No. DE-SC0020633.
J.N. is partially supported by the U.S. Department of Energy, Office of Science, Office for Nuclear Physics
under Award No. DE-SC0021301. D.A. is supported by the U.S. Department of Energy, Office of Science, Office for Nuclear Physics under Award No. DE-SC0013470.

\end{acknowledgments}

%

\appendix

\onecolumngrid

\section{Supplemental Material}
In this Supplemental Material we give the details about the calculations performed in the main text. The hydrodynamic equations solved in this work, and also the sufficient and necessary conditions for nonlinear causality derived in \cite{Bemfica:2020xym} and used in this work, are presented in Section \ref{sec:I}. The analysis concerning the characteristic velocities can also be found in Section \ref{sec:I}. The corresponding causality analysis in small systems (proton-nucleus collisions) simulations can be found in Section \ref{sec:II}. Details concerning the effects on final observables can be found in Section \ref{sec:III}.
\subsection{Hydrodynamic equations of motion, causality conditions, and characteristic velocities}\label{sec:I}
\begin{figure*}
    \centering
     \includegraphics[keepaspectratio, width=0.39\linewidth]{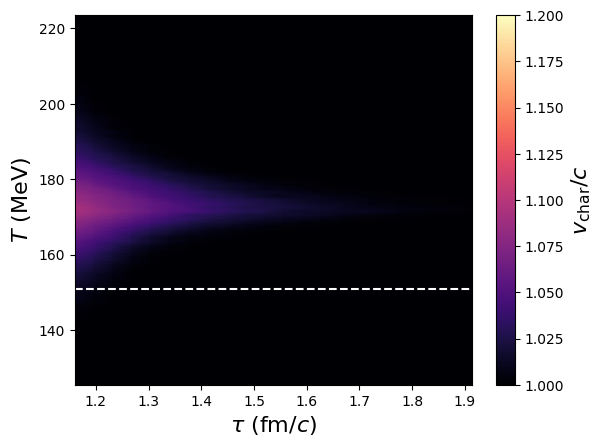}%
     \includegraphics[keepaspectratio, width=0.39\linewidth]{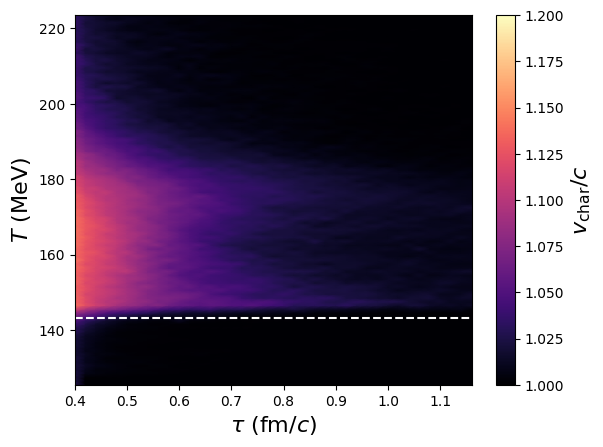}\\
      \includegraphics[keepaspectratio, width=0.39\linewidth]{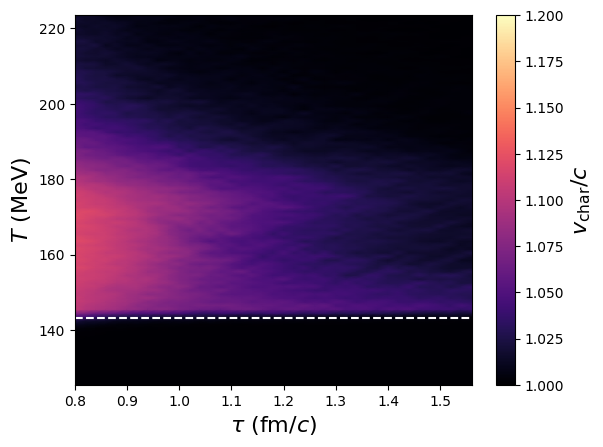}
     \includegraphics[keepaspectratio, width=0.39\linewidth]{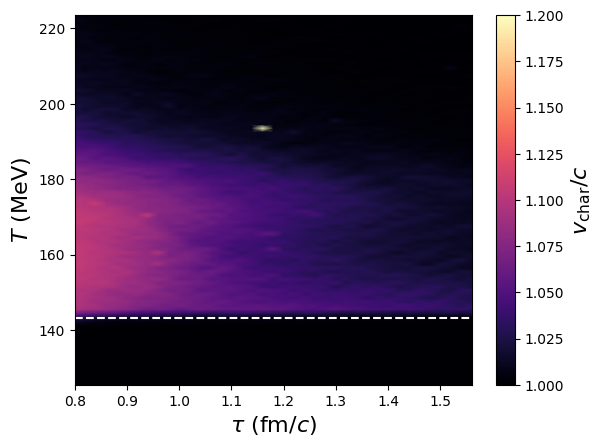}
    \caption{Characteristic Velocity, Pb+Pb: Top left: TFV. Top right: IKM, no Kompost. Bottom left: IKM, FS Kompost, Bottom right: IKM, EKT Kompost.  See text for discussion.}
    \label{Fig:velocity}
\end{figure*}
In this  work, we consider the DNMR-type hydrodynamics equations \cite{Denicol:2012cn} (at zero chemical potential) where the energy-momentum is written in the Landau hydrodynamic frame \cite{landaulifshitzfluids} as $T_{\mu\nu} = \varepsilon\, u_\mu u_\nu 
+( P+\Pi)\Delta_{\mu\nu}+\pi_{\mu\nu}$, where $\varepsilon$ is the Lorentz scalar energy density, $P$ is the thermodynamical pressure defined by the equation of state, $u_\mu$ is the 4-flow velocity, $\Delta_{\mu\nu} = g_{\mu\nu} + u_\mu u_\nu$ (with $g_{\mu\nu}$ being the Minkowski metric), $\Pi$ is the bulk scalar, and $\pi_{\mu\nu}$ is the shear-stress tensor. The hydrodynamic equations of motion stem from  energy-momentum conservation, $\nabla_\mu T^{\mu\nu}=0$, supplemented by the following equations for the dissipative quantities $\{\pi_{\mu\nu},\Pi\}$ 
\bml
\label{supplemental}
\bea
\tau_\Pi u^\mu \nabla_\mu \Pi + \Pi &=& -\zeta \nabla_\mu u^\mu -\delta_{\Pi\Pi}\Pi \nabla_\mu u^\mu -\lambda_{\Pi \pi} \pi^{\mu\nu}\sigma_{\mu\nu},\label{bulk}\\
 \tau_\pi \Delta^{\mu\nu}_{\alpha\beta}u^\lambda \nabla_\lambda \pi^{\alpha\beta} + \pi^{\mu\nu} &=& -2 \eta \sigma^{\mu\nu} -\delta_{\pi \pi} \pi^{\mu\nu}\nabla_\alpha u^\alpha  -\tau_{\pi\pi}\pi_\alpha^{\langle \mu}\sigma^{\nu\rangle\alpha} - \lambda_{\pi\Pi}\Pi \sigma^{\mu\nu},
\eea
\eml
where $\sigma_{\mu\nu}$ is the shear tensor. The particular form of parametrization for the shear and bulk viscosity transport coefficients, $\{\eta,\zeta\}$, and all other second order transport coefficient $\{\tau_\pi, \tau_\Pi,\delta_{\Pi\Pi},\lambda_{\Pi \pi},\delta_{\pi \pi},\tau_{\pi\pi},\lambda_{\pi\Pi}\}$ was based on the parametrization used in \cite{Moreland:2018gsh, Bernhard:2019bmu}, \cite{Schenke:2012hg, Schenke:2012wb}.
\subsubsection{Nonlinear causality conditions}
The nonlinear causality conditions presented in \cite{Bemfica:2020xym} were obtained by studying the characteristic velocities of the corresponding nonlinear set of partial differential equations. The constraints of \cite{Bemfica:2020xym} can be divided into two sets of conditions. The first set defines \emph{necessary} conditions that must be satisfied otherwise causality is certainly violated. These necessary conditions are given by \cite{Bemfica:2020xym}:

\bml
\label{necessary_conditions}
\bea
&&(2 \eta+\lambda_{\pi\Pi}\Pi)-\frac{1}{2}\tau_{\pi\pi}|\Lambda_1| \geq 0\\
&&\varepsilon+P+\Pi-\frac{1}{2\tau_\pi}(2 \eta+\lambda_{\pi\Pi}\Pi)-\frac{\tau_{\pi\pi}}{4\tau_\pi}\Lambda_3\ge 0,\label{n12}\\
&&
\frac{1}{2\tau_\pi}(2 \eta+\lambda_{\pi\Pi}\Pi)+\frac{\tau_{\pi\pi}}{4\tau_\pi}\left (\Lambda_a +\Lambda_d\right ) \geq 0,\quad a\ne d,
\label{new_necessary_condition}
\\
&&\varepsilon+P+\Pi+\Lambda_a-\frac{1}{2\tau_\pi}(2 \eta+\lambda_{\pi\Pi}\Pi)-\frac{\tau_{\pi\pi}}{4\tau_\pi}\left (\Lambda_d +\Lambda_a\right )\ge0,\quad a\ne d\label{n11}\\
&& \frac{1}{2\tau_\pi}(2 \eta+\lambda_{\pi\Pi}\Pi)+\frac{\tau_{\pi\pi}}{2\tau_\pi}\Lambda_d
+\frac{1}{6\tau_\pi}[2 \eta+\lambda_{\pi\Pi}\Pi+(6\delta_{\pi\pi}-\tau_{\pi\pi})\Lambda_d]
\nonumber \\
&& +\frac{\zeta+\delta_{\Pi\Pi}\Pi+\lambda_{\Pi \pi}\Lambda_d}{\tau_\Pi}+(\varepsilon+P+\Pi+\Lambda_d)c_s^2 \geq 0,
\label{n1}\\
&& \varepsilon+P+\Pi+\Lambda_d -\frac{1}{2\tau_\pi}(2 \eta+\lambda_{\pi\Pi}\Pi)-\frac{\tau_{\pi\pi}}{2\tau_\pi}\Lambda_d
-\frac{1}{6\tau_\pi}[2 \eta+\lambda_{\pi\Pi}\Pi+(6\delta_{\pi\pi}-\tau_{\pi\pi})\Lambda_d]
\nonumber\\
&&-\frac{\zeta+\delta_{\Pi\Pi}\Pi+\lambda_{\Pi \pi}\Lambda_d}{\tau_\Pi}-(\varepsilon+P+\Pi+\Lambda_d)c_s^2 \geq 0,
\label{n2}\eea
\eml
where \eqref{new_necessary_condition}-\eqref{n2} must hold for $a,d=1, 2,3$, and $c_s^2 = dP/d\varepsilon$ is the speed of sound squared.

On the other hand, Ref.\ \cite{Bemfica:2020xym} also derived \emph{sufficient} conditions for causality, i.e., conditions which, if satisfied, for sure guarantee that the evolution is causal. These sufficient conditions are given by \cite{Bemfica:2020xym}:
 \bml
\label{conditions}
\bea
&&(\varepsilon+P+\Pi-|\Lambda_1|)-\frac{1}{2\tau_\pi}(2 \eta+\lambda_{\pi\Pi}\Pi)-\frac{\tau_{\pi\pi}}{2\tau_\pi}\Lambda_3\ge 0,\label{cond1-1}\\
&&(2 \eta+\lambda_{\pi\Pi}\Pi)-\tau_{\pi\pi}|\Lambda_1|>0,\label{cond1-2}\\
&&\tau_{\pi\pi}\le 6\delta_{\pi\pi},\label{cond4a}\\
&&\frac{\lambda_{\Pi \pi} }{\tau_\Pi}+c_s^2-\frac{\tau_{\pi\pi}}{12\tau_\pi}\ge 0,\label{cond4b}\\
&&\frac{1}{3\tau_\pi}[4 \eta+2\lambda_{\pi\Pi}\Pi+(3\delta_{\pi\pi}+\tau_{\pi\pi})\Lambda_3]+\frac{\zeta+\delta_{\Pi\Pi}\Pi+\lambda_{\Pi \pi}\Lambda_3}{\tau_\Pi}+|\Lambda_1|+\Lambda_3 c_s^2 \nonumber\\
&&+\frac{\frac{12\delta_{\pi\pi}-\tau_{\pi\pi}}{12\tau_\pi}\left (\frac{\lambda_{\Pi \pi} }{\tau_\Pi}+c_s^2-\frac{\tau_{\pi\pi}}{12\tau_\pi}\right )(\Lambda_3+|\Lambda_1|)^2}{\varepsilon+P+\Pi-|\Lambda_1|-\frac{1}{2\tau_\pi}(2 \eta+\lambda_{\pi\Pi}\Pi)-\frac{\tau_{\pi\pi}}{2\tau_\pi}\Lambda_3}\le (\varepsilon+P+\Pi)(1-c_s^2),\label{cond5}\\
&&\frac{1}{6\tau_\pi}[2 \eta+\lambda_{\pi\Pi}\Pi+(\tau_{\pi\pi}-6\delta_{\pi\pi})|\Lambda_1|]+\frac{\zeta+\delta_{\Pi\Pi}\Pi-\lambda_{\Pi \pi}|\Lambda_1|}{\tau_\Pi}+(\varepsilon+P+\Pi-|\Lambda_1|)c_s^2 \ge 0,\label{cond7}\\
&&1\ge\frac{\frac{12\delta_{\pi\pi}-\tau_{\pi\pi}}{12\tau_\pi}\left (\frac{\lambda_{\Pi \pi} }{\tau_\Pi}+c_s^2-\frac{\tau_{\pi\pi}}{12\tau_\pi}\right )(\Lambda_3+|\Lambda_1|)^2}{\left [\frac{1}{2\tau_\pi}(2 \eta+\lambda_{\pi\Pi}\Pi)-\frac{\tau_{\pi\pi}}{2\tau_\pi}|\Lambda_1|\right ]^2}\label{cond6}\\
&&\frac{1}{3\tau_\pi}[4 \eta+2\lambda_{\pi\Pi}\Pi-(3\delta_{\pi\pi}+\tau_{\pi\pi})|\Lambda_1|]+\frac{\zeta+\delta_{\Pi\Pi}\Pi-\lambda_{\Pi \pi}|\Lambda_1|}{\tau_\Pi}+(\varepsilon+P+\Pi-|\Lambda_1|)c_s^2\nonumber\\
&&\ge \frac{(\varepsilon+P+\Pi+\Lambda_2)(\varepsilon+P+\Pi+\Lambda_3)}{3(\varepsilon+P+\Pi-|\Lambda_1|)}\left \{1+\frac{2\left [\frac{1}{2\tau_\pi}(2 \eta+\lambda_{\pi\Pi}\Pi)+\frac{\tau_{\pi\pi}}{2\tau_\pi}\Lambda_3\right ]}{\varepsilon+P+\Pi-|\Lambda_1|}\right \}.\label{cond8}
\eea
\eml
Above, $\Lambda_1$, $\Lambda_2$, and $\Lambda_3$ are eigenvalues of $\pi_{\mu}^{\nu}$ which obey $\Lambda_1 + \Lambda_2 + \Lambda_3 = 0$ (due to the fact that the shear-stress tensor is traceless) and $\Lambda_1 \leq \Lambda_2 \leq \Lambda_3$ with $\Lambda_1 \leq 0 \leq \Lambda_3$. The set of necessary conditions and, the set of sufficient conditions, given  above were the ones used in the main part of the text. We refer the reader to \cite{Bemfica:2020xym} for the derivation of the conditions shown above.  


\subsubsection{Characteristic velocities}

\begin{figure*}
    \centering
    \includegraphics[keepaspectratio, width=0.39\linewidth]{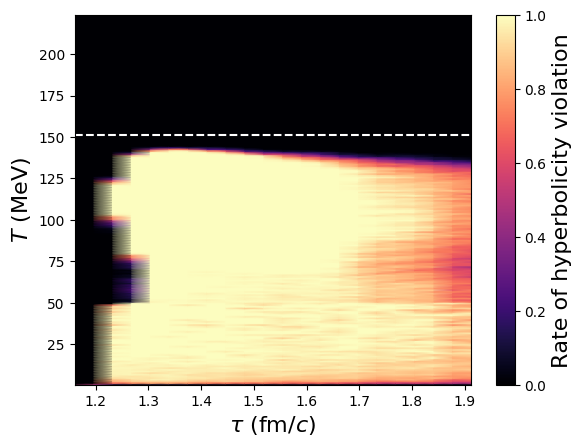}
     \includegraphics[keepaspectratio, width=0.39\linewidth]{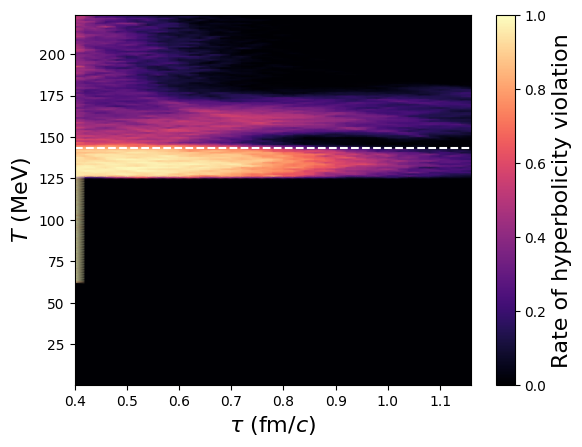}\\
    \includegraphics[keepaspectratio, width=0.39\linewidth]{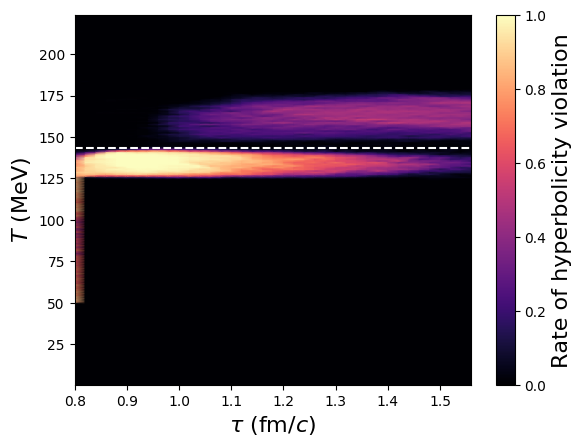}
    \includegraphics[keepaspectratio, width=0.39\linewidth]{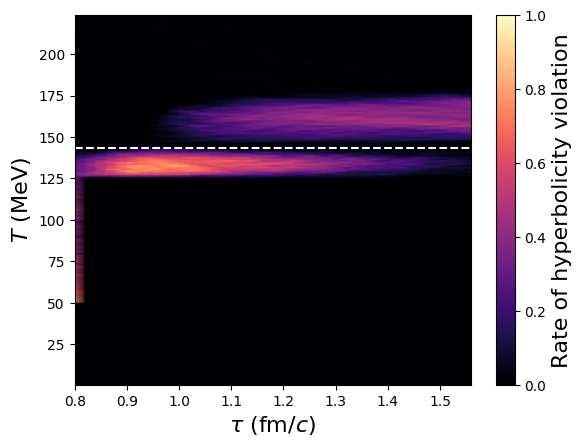}
    \caption{Hyperbolicity violations in Pb+Pb system: Top left: TFV. Top right: IKM, no Kompost. Bottom left: IKM, FS Kompost, Bottom right: IKM, EKT Kompost.  See text for discussion.}
    \label{Fig:hyperbolicity}
\end{figure*}
Causality can be investigated by determining the characteristic manifolds associated with a given system of partial differential equations. The corresponding characteristic surfaces $\{\Phi(x)=0\}$ are determined by the principal part of the equations of motion by solving the characteristic equation for the normal 4-vector $\xi_\alpha = \nabla_\alpha \Phi$ \cite{Bemfica:2020xym}. The system is causal if, for any spatial components $\xi_i$, the roots $\xi_0 = \xi_0(\xi_i)$ of the characteristic equation are real and the 4-vector $\xi_\alpha = (\xi_0,\xi_i)$ is spacelike or lightlike. This procedure determines the so-called characteristic velocities, which must be real-valued quantities that cannot exceed unity (in natural units where $\hbar = c=k_B=1$) in order to prevent superluminal propagation, i.e., acausal behavior. In our work, we show that there are cells in the hydrodynamic evolution where the characteristic velocities are larger than unity (up to $15\%$ larger than the speed of light) and there are also cells where the characteristic velocities are not real numbers (which signals hyperbolicity violation). 

In the Supplemental Material of Ref.\ \cite{Bemfica:2020xym}, the full expressions and inequalities involving the characteristic velocities were given. In the numerical analysis done in our work, we use the following expressions to quantify the characteristic velocities of the fluid dynamical evolution: Eqs.~(S7-S11) in  \cite{Bemfica:2020xym}. The definition of the $\mathbf{\mathfrak{g}}_a$  immediately following Eq.(S3) in \cite{Bemfica:2020xym} is
\[\mathfrak{g}_{a}=\frac{2(2 \eta+\lambda_{\pi\Pi}\Pi)+\tau_{\pi\pi}\Lambda_a}{4 \rho \tau_\pi}.\]
Setting each of the factors $m_a, m_b$ equal to zero, we obtain the roots
\be
\label{ma}
 k=\frac{\frac{1}{2\tau_\pi}(2 \eta+\lambda_{\pi\Pi}\Pi)+\frac{\tau_{\pi\pi}}{4\tau_\pi}\left (\Lambda_a +\Lambda_d\right )}{\varepsilon+P+\Pi+\Lambda_a},\quad a\ne d.
 \ee
Causality is violated if $k<0$, leading to condition (4c), of if $k>1$, leading to condition (4d).
The remaining root above is obtained when the term in brackets vanishes, giving
\bea
\label{nf2}
&& k=\frac{\frac{1}{2\tau_\pi}(2 \eta+\lambda_{\pi\Pi}\Pi)+\frac{\tau_{\pi\pi}}{2\tau_\pi}\Lambda_d}{\varepsilon+P+\Pi+\Lambda_d}\nonumber\\
&&+\frac{\bigg \{\frac{1}{6\tau_\pi}[2 \eta+\lambda_{\pi\Pi}\Pi+(6\delta_{\pi\pi}-\tau_{\pi\pi})\Lambda_d]+\frac{\zeta+\delta_{\Pi\Pi}\Pi+\lambda_{\Pi \pi}\Lambda_d}{\tau_\Pi}+(\rho+\Lambda_d)c_s^2\bigg \}}{\varepsilon+P+\Pi+\Lambda_d}.
\eea
Causality is violated if $k< 0$, leading to (4e), or if $k> 1$, leading to (4f).

\subsubsection{Comparison of superluminal propagation and hyperbolicity violations in the TFV and IKM frameworks}
In Fig.~\ref{Fig:velocity} and Fig.~\ref{Fig:hyperbolicity}, we show in detail how to quantify the degree of causality violations in our simulations.  Fig.~\ref{Fig:velocity} shows the distribution of average superluminal propagation in fluid cells as a function of the proper time $\tau$ and temperature $T$.  The superluminal (SL) distribution is computed by
\begin{equation}
  \rho_\mathrm{SL}(\tau, T) \equiv \frac{1}{N_\mathrm{cells}} \sum_{i = 0}^{N_\mathrm{cells}} \delta v^{(i)}_\mathrm{char} \theta\l( \delta v^{(i)}_\mathrm{char} \r),\label{SLdist}
\end{equation}
where $N_\mathrm{cells}(\tau,T)$ is the total number of cells for a given $\tau$ and $T$, $\delta v^{(i)}_\mathrm{char} \equiv v^{(i)}_\mathrm{char} - 1$ is the maximum superluminal violation in the $i$th cell, and $\theta(x)$ is the Heaviside theta function.  Note that, by construction, all cells contribute to the denominator of \eqref{SLdist}, but only those with superluminal violations contribute to the numerator.

We find that the rate of superluminal violation in the TFV and IKM frameworks reaches up to the scale of $v_\mathrm{char}/c \sim 15\%$, with the most prevalent violations occuring at early proper times and at temperatures above freeze out.  We note that the TFV framework encounters smaller violations which dissipate more quickly, whereas the IKM frameworks encounters larger violations which persist for longer.  Additionally, we observe that including pre-equilibrium evolution in the IKM scenarios helps to reduce the magnitude of the violations while also extending their duration in $\tau$.

In Fig.~\ref{Fig:hyperbolicity} we present the rate of hyperbolicity (HB) violations in our simulations.  The hyperbolicity violations are defined by
\begin{equation}
  \rho_\mathrm{HB}(\tau, T) \equiv \frac{N_\mathrm{HB}}{N_\mathrm{cells}},\label{HBdist}
\end{equation}
where $N_\mathrm{HB}(\tau,T)$ is the number of cells at a given $\tau$ and $T$ where the minimum characteristic squared velocity (either $\mathbf{\mathfrak{g}}_a$ or $k$ above) is negative, indicating a failure of hyperbolicity in the equations of motion.  A value of $\rho_\mathrm{HB} = 1$ thus implies that all cells at a given $\tau$ and $T$ are failing to evolve in a hyperbolic fashion.

We find that the hyperbolicity violations are most prevalent below freeze out, although a smaller fraction occurs also above freeze out for some scenarios.  The TFV framework exhibits by far the most severe violations in this respect.  We note also a possible complementarity between the distributions of superluminal violations and hyperbolicity violations shown in Figs.~\ref{Fig:velocity} and \ref{Fig:hyperbolicity}: simulations which fare better in superluminal violations seem to exhibit stronger hyperbolicity violations, and vice versa.  This may indicate that some degree of non-hyperbolic evolution is required in order for simulations to eliminate superluminal propagation in the initial timesteps.  We defer further discussion of this possibility to future work.

\subsection{Causality analysis for small systems}\label{sec:II}
\begin{figure}
    \centering
    \includegraphics[keepaspectratio, width=0.245\linewidth]{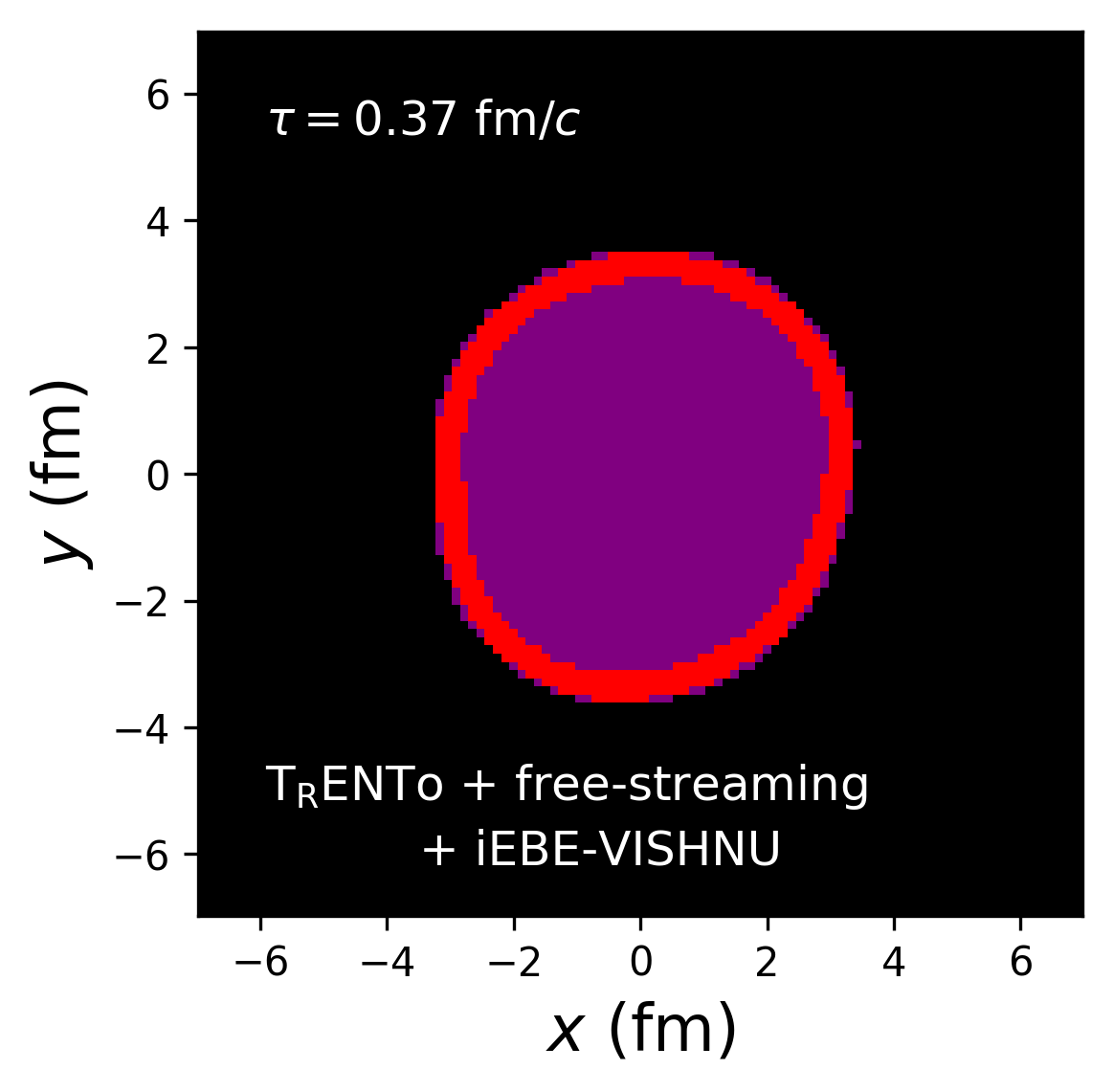}%
     \includegraphics[keepaspectratio, width=0.33\linewidth]{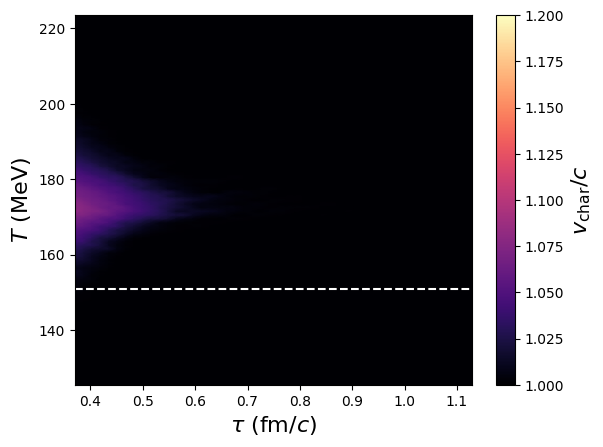}%
    \includegraphics[keepaspectratio, width=0.33\linewidth]{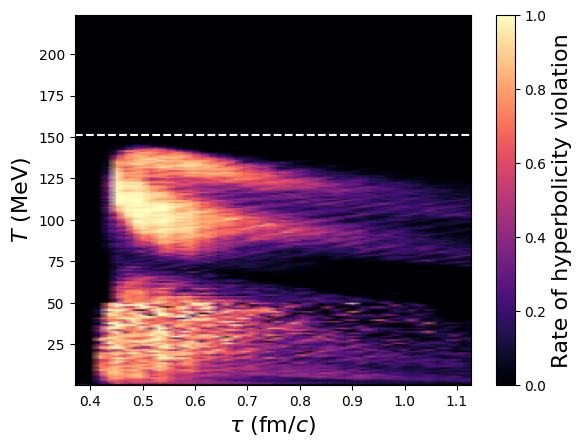}\\
    \includegraphics[keepaspectratio, width=0.25\linewidth]{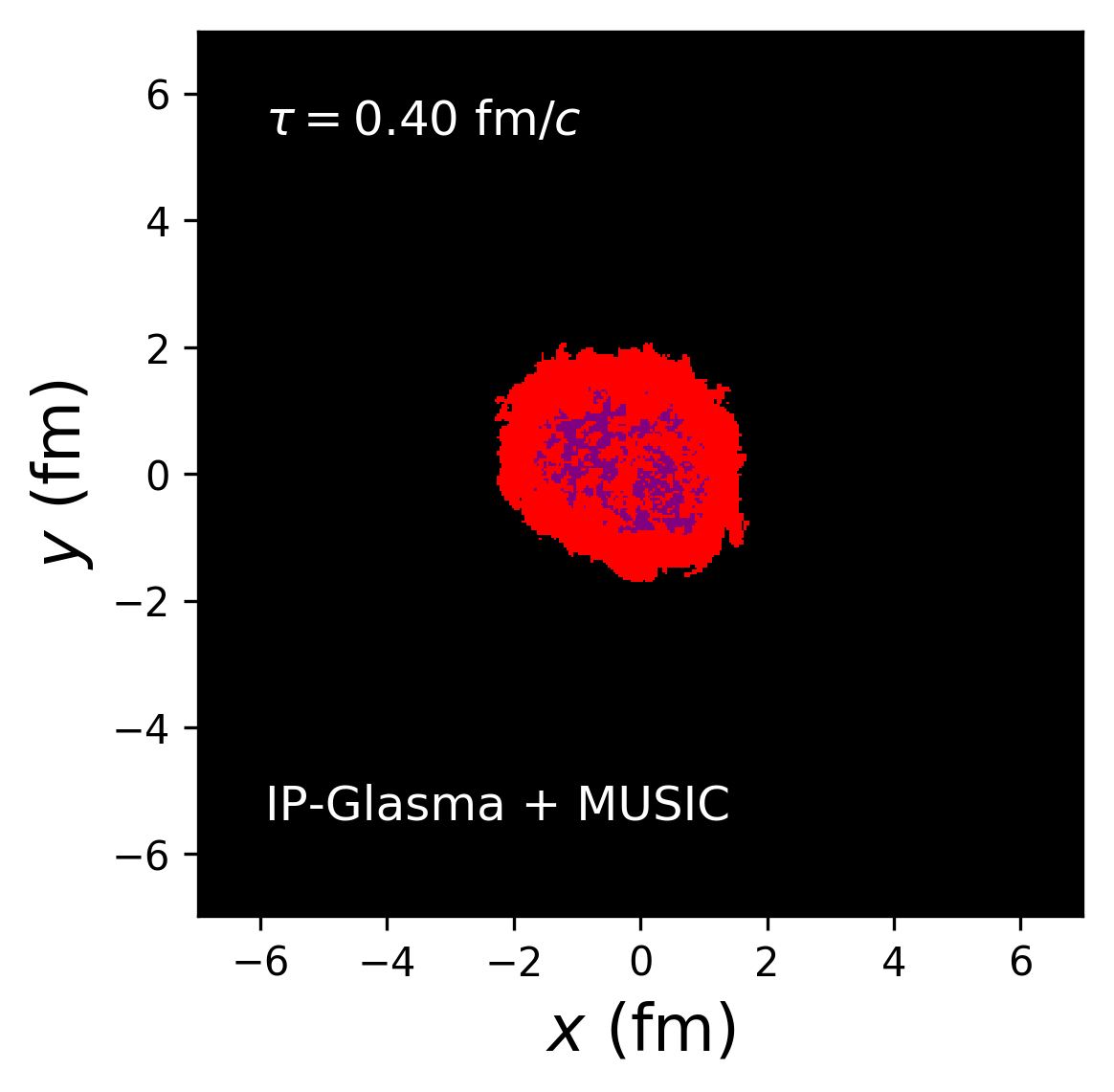}
     \includegraphics[keepaspectratio, width=0.32\linewidth]{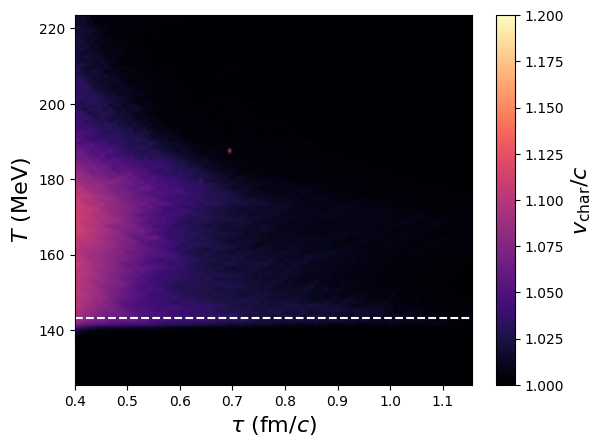}%
    \includegraphics[keepaspectratio, width=0.32\linewidth]{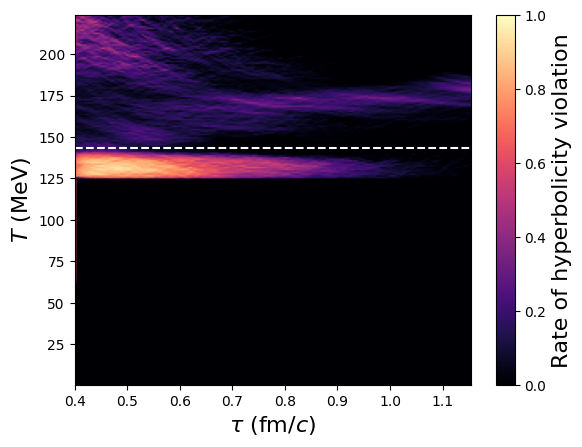}
    \caption{Left: Causality violations in the initial timestep for the VISHNU framework (left) and the MUSIC framework without KoMPoST evolution (right) in p+Pb at $\sqrt{s}=2.76$ TeV.  Note that the lefthand panel includes some pre-equilibrium free-streaming, whereas the righthand panel does not. Middle: Characteristic Velocity. Righ: hyperbolicity violations }
    \label{Fig:smallsys}
\end{figure}
Fig.~\ref{Fig:smallsys} demonstrates that causality violation is equally or more problematic in small systems such as p+Pb.  However, the fact that the violations are not much more prevalent in the VISHNU framework event leads one to be cautiously optimistic that hydrodynamics could eventually be extended successfully even to small systems, albeit only in the correct regime of model parameter space.
Fig.~\ref{Fig:smallsys} illustrate the characteristic velocity and hyperbolicity violations for p+Pb simulations using both frameworks discussed in the main text.  The behavior observed in Pb+Pb collisions in the IKM framework agrees qualitatively with that observed here in small systems: superluminal propagation is again seen in excess of 10\% over the speed of light.  Some hyperbolicity violations are noted above the freeze-out temperature, but the highest rate of these violations occur below freeze-out and above the regulator cutoff (below which the rate of violations drop rapidly to zero).  The TFV framework performs somewhat better with respect to the superluminal violations but fails severely with respect to hyperbolicity violations, reaching nearly 100\% for certain proper times and temperatures below freeze out.  In essentially both cases, the violations die away within roughly the first 1 fm$/c$ of the hydrodynamic evolution.

\subsection{Effects on final observables}\label{sec:III}

Table \ref{table} shows the effects of including or excluding acausal/indeterminate fluid cells in the evaluation of the initial and final $\epsilon_{2,p}$ and $\epsilon_{2,x}$.  As is standard practice, the anisotropies represent integrals over the full transverse plane, implying that cells with $\varepsilon \leq \varepsilon_{FO}$ are also included when evaluating 

\begin{table*}[]
\begin{tabular}{|l|r|r|r|r|r|r|r|r|}
\hline
 & \multicolumn{2}{c|}{Initial}  & \multicolumn{2}{|c|}{Final} & \multicolumn{2}{|c|}{Initial} & \multicolumn{2}{|c|}{Final} \\
\hline
 & $\epsilon_{2,x}$ [all] & $\epsilon_{2,x}$ [causal] & $\epsilon_{2,x}$ [all] & $\epsilon_{2,x}$ [causal] & $\epsilon_{2,p}$ [all] & $\epsilon_{2,p}$ [causal] & $\epsilon_{2,p}$ [all] & $\epsilon_{2,p}$ [causal] \\ \hline
VISHNU & 0.0396 & 0.0515 & 0.0218 & 0.0347 & 0.00281 & 0.0238 & 0.0277 & 0.0348 \\ \hline
MUSIC (EKT pre-eq.) & 0.101 & 0.119 & 0.406 & 0.141 & 0.0177 & 0.0372 & 0.0620 & 0.0731 \\ \hline
MUSIC (FS pre-eq.) & 0.101 & 0.141 & 0.0461 & 0.161 & 0.0156 & 0.0253 & 0.0630 & 0.0918 \\ \hline
MUSIC (no pre-eq.) & 0.0997 & 0.120 & 0.0335 & 0.156 & 0.0074 & 0.0233 & 0.0528 & 0.0882 \\ \hline
\end{tabular}
\caption{A summary of the effects of acausal/indeterminate cells on the initial and final values of $\epsilon_{2,p}$ and $\epsilon_{2,x}$.  The initial values are averaged over the early stages of each system until at least half of the fluid cells are causal; the final values are taken from the latest freeze-out time, using either all cells at the final timestep, or using only cells which are causal at the final timestep. \label{table}}
\end{table*}

\begin{align}
\epsilon_{2,p}&=\sqrt{\l( \l< T^{xx}-T^{yy}\r>_1^2 + \l< 2T^{xy} \r>_1^2 \r) / \l< T^{xx}+T^{yy} \r>_1^2} \label{e2pdef},\\ 
\epsilon_{2,x}&= \sqrt{\l( \l< x^2-y^2\r>_{e \gamma}^2 + \l< 2 x y \r>_{e \gamma}^2 \r) / \l< x^2+y^2 \r>_{e \gamma}^2},
\label{e2xdef}
\end{align}
where $\l< f(x,y) \r>_w= \l.\int dx\, dy\, w(x,y) f(x,y) \r/ \int dx\, dy\, w(x,y)$ and $\gamma = \sqrt{1-u_x^2-u_y^2}$.

To minimize numerical instabilities, we evaluate the initial anisotropies by first averaging the numerator and denominator appearing in \eqref{e2pdef} and \eqref{e2xdef} separately over all times up to the point at which exactly half of the fluid cells are causal.

\end{document}